\documentclass[12]{article}

\usepackage{cite}
\usepackage{times}
\usepackage{fullpage}
\usepackage{lipsum}
\usepackage{amsmath}
\usepackage{amsfonts}
\usepackage{amssymb}
\usepackage{latexsym}
\usepackage{dsfont,eucal,mathrsfs}
\usepackage{graphicx}
\usepackage{subcaption}
\usepackage{bm}
\usepackage{ragged2e}
\usepackage{hyperref}
\DeclareCaptionJustification{justified}{\justifying}
\usepackage{todonotes}
\usepackage{changes}

\definechangesauthor[name={Methods}, color=blue]{Methods}

\newcommand{\spacing}[1]{\renewcommand{\baselinestretch}{#1}\large\normalsize}
\spacing{2}

\makeatletter
\renewcommand{\maketitle}{\bgroup\setlength{\parindent}{0pt}
\begin{flushleft}
  {\Large\bfseries\noindent\sloppy \textsf{\@title} \par}

  \@author
\end{flushleft}\egroup
}
\makeatother

\newenvironment{affiliations}{%
    \setcounter{enumi}{1}%
    \setlength{\parindent}{0in}%
    \slshape\sloppy%
    \begin{list}{\upshape$^{\arabic{enumi}}$}{%
        \usecounter{enumi}%
        \setlength{\leftmargin}{0in}%
        \setlength{\topsep}{0in}%
        \setlength{\labelsep}{0in}%
        \setlength{\labelwidth}{0in}%
        \setlength{\listparindent}{0in}%
        \setlength{\itemsep}{0ex}%
        \setlength{\parsep}{0in}%
        }
    }{\end{list}\par\vspace{12pt}}

\usepackage{mwe}

\renewcommand{\mathbf}{\boldsymbol}
\renewcommand{\mathcal}{\mathscr}

\renewenvironment{abstract}{%
    \setlength{\parindent}{0in}%
    \setlength{\parskip}{0in}%
    \bfseries%
    }{\par\vspace{-6pt}}

\title{Consequences of delays and imperfect implementation
of isolation in epidemic control}

\author
{Lai-Sang Young,$^{1 \ast}$ Stefan Ruschel,$^{2}$ Serhiy Yanchuk,$^2$ Tiago Pereira$^{3,4}$}

\date{}

\begin{document}

\maketitle

\begin{affiliations}
 \item Courant Institute of Mathematical Sciences, New
York University
\item Institut f\"ur Mathematik, Technische Universit\"at
Berlin
\item Instituto de Ciencias Matem\'aticas e Computa\c{c}\~ao, Universidade de S\~ao Paulo, S\~ao Carlos, Brazil
\item Department of Mathematics, Imperial College London, London SW7 2AZ, UK

 \end{affiliations}

\begin{abstract}
%
{
For centuries isolation has been the main control strategy of unforeseen epidemic outbreaks. When implemented in full and without delay, isolation is very effective. However, flawless implementation is seldom feasible in practice. We present an epidemic model called SIQ with an isolation protocol, focusing on the consequences of delays and incomplete identification of infected hosts. The continuum limit of this  model is a system of Delay Differential Equations, the analysis of which reveals clearly the dependence of epidemic evolution on model parameters including  disease reproductive  number, isolation probability, speed of identification of infected hosts and recovery rates. Our model offers estimates on minimum response capabilities needed to  curb outbreaks, and predictions of endemic states when containment fails. Critical response capability is expressed explicitly in terms of parameters that are easy to obtain, to assist in the evaluation of funding priorities involving preparedness and epidemics management.
}
\end{abstract}


Roughly 400 infectious diseases have been identified since 1940.  New pathogens are emerging at higher rates, despite the increase in awareness and  vigilance.   A grave public health concern  is when and how the next outbreak will occur \cite{Woolhouse2016}; threats of imminent global outbreaks are real \cite{tian2015}.  The 1917 Spanish influenza, which killed 50 million people, was the worst-ever pandemic on record -- and that was back at a time when travel by ship was the fastest means of transportation around the globe.  In today's tightly connected world, an epidemic can potentially travel at jet-speed. Indeed, Swine flu was first detected in April of 2009 in Mexico and within a week it turned up in London. 

In spite of advances in medicine, there is still no known vaccine for many infectious diseases, and even when a vaccine is known, it is generally impractical to maintain an adequate stockpile.
Preventive measures may fall out of fashion when the disease is absent for long periods, only to hit harder when it returns due to the lower level of immunity.
A standard  recommendation of health organizations in the face of  an outbreak is the isolation of infected individuals \cite{CDC_legal,Siegel2007,centers2014}. This control strategy dates back centuries and  its usefulness has not diminished with time, as evidenced in the recent outbreaks of SARS in Asia and Ebola in West Africa. Research has progressed in the meantime;  mathematical models to understand the impact and effectiveness of isolation and quarantine protocols have been proposed and analyzed \cite{Zuzek2015,Grigoras2016,Reppas2010,Day2006}; it has been shown that the effectiveness of isolation protocols can be increased by contact tracing and isolation of individuals prior to symptoms \cite{Fraser2004,Peak2017}.

Perfect implementation of isolation, however, can never be achieved.  Among the obstacles are the following: Physical facilities to house the isolated population, and medical personnel to care for them, have to be available \cite{Peak2017,kucharski2015}. Even then, the identification of infected hosts, a crucial first step in any isolation strategy,  can be a formidable task.  At early stages of an outbreak, the general population -- even local health  authorities -- may fail to recognize the symptoms of an infectious disease (which are often ambiguous), or they may not be cognizant of  the potential dangers of runaway exponential growth of  infectious diseases \cite{Day2006}. For reasons of their own, such as distrust of authorities, infected individuals may choose not to seek medical attention. The cost of preparedness and surveillance, and the maintenance of adequate infrastructure and personnel that can be deployed at a moment's notice, can be prohibitively high \cite{Herstein2017}.

A case in point was the initial less-than-adequate emergency response in the devastating Ebola epidemics in West Africa. It was not until Oct 2014 that billions of dollars were allocated for training of personnel, construction and  maintenance of holding centers and treatment units.  Once the response capability improved, and delays in identifying infected individuals were shortened, the spreading of the disease was more effectively curtailed \cite{kucharski2015,team2014}. Detrimental effects of delay were also reported during the SARS outbreak, which became more effectively controlled only when the time between onset of symptoms and isolation was shortened \cite{Donnelly2003,kucharski2015}.
Though recent studies have used linear analysis [6,16] and simulations [9]
to investigate the role of isolation, a general comprehensive theory of isolation as a means of epidemic control that focuses on consequences of imperfect implementation is lacking, as are studies of the nonlinear effects of isolation protocols.

This paper addresses fundamental issues pertaining to imperfections in the implementation of isolation strategies as a means of disease control. First and foremost is the minimum response capability required to contain an outbreak. We determine the {\it isolation probability}, the critical fraction of the infected population that must be isolated, and the {\it identification time}, how quickly local authorities must act to prevent a full-blown outbreak. Our model predicts the nature of the endemic state that follows should the response be inadequate, and the extent to which
a stronger response can ameliorate the severity of an outbreak.
It offers crucial information to health authorities such as estimates on the fraction of the population that can be expected to fall ill, and answers theoretical questions such as the potential benefits of longer durations of isolation. 

These and other questions of a highly nonlinear nature are best tackled using dynamical systems techniques \cite{Guckenheimer2002,Lu2013}. In the present work, we analyzed the continuum limit of  an isolation strategy. Under some assumptions on identification and isolation times, we obtained a system of Delay Differential Equations (DDE) describing the time course of an infection following an outbreak. These equations are amenable to mathematical analysis \cite{Hale1993,YanchukGiacomelli2017}, enabling us  to understand rigorously many of the issues above. Our model is complex enough to capture key ideas, and simple enough to permit the representation of relevant quantities explicitly in terms of disease and control parameters. We view these as among our model's greatest strengths.

\section*{Results}

\begin{figure*}
\centering
\includegraphics[width=0.6\linewidth]{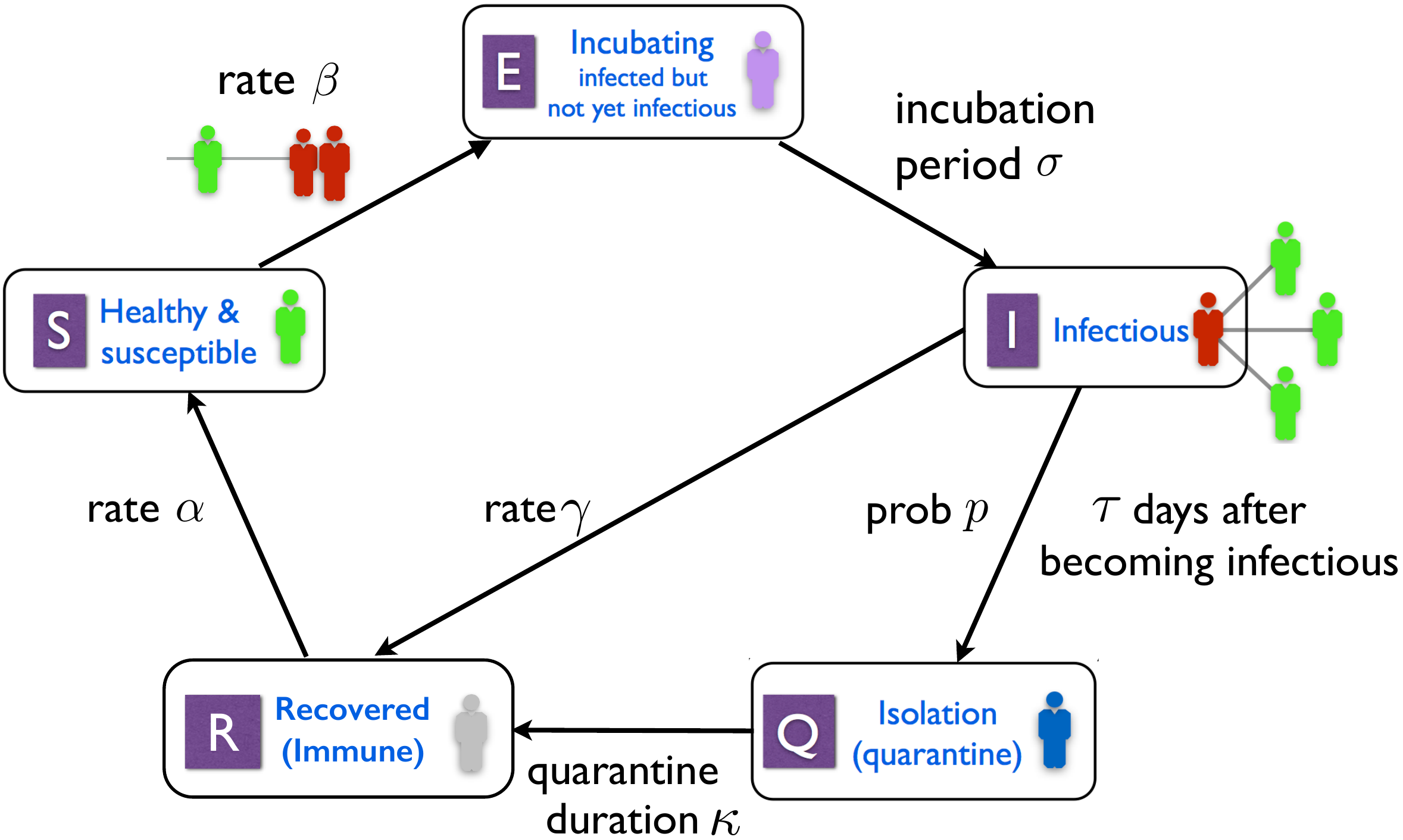}%
\caption{\label{fig:model}
%
The SIQ model. Healthy individuals in contact with an infected node become infected at rate $\beta$. Once infected, a node incubates the disease for $\sigma$ units of time, during which it is not symptomatic  and cannot transmit the disease.  At the end of incubation, these nodes become infectious. Infectious nodes recover on their own at rate $\gamma$ gaining immunity against the disease. Immunity is lost at a rate $\alpha$. Those infectious nodes that are not recovered after $\tau$ units of time are, with probability $p$, identified and isolated. They remain in isolation for a duration of $\kappa$ at the end of which they are released and rejoin the healthy group.
%
}
\end{figure*}

\subsection*{SIQ: an epidemic model with (partial) isolation}
Consider, to begin with, a random network of $N$ nodes;  each node represents a host, and nodes that are linked by edges are neighbors.  The mean degree at a node is given by $\langle k \rangle$, and an important parameter for us is the density of  connections  $ m = \langle k \rangle  / N$ at each node. In the absence of any control mechanism, the situation is as follows. Each host is in one of four discrete states: healthy and susceptible ($S$), incubating ($E$), infectious ($I$), and recovered (R). Infectious hosts  infect their neighbors at rate $\beta$; infected nodes incubate the disease for a period of $\sigma$ units of time during  which they are assumed to be neither symptomatic nor infectious. At the end of the incubation period they become infectious. Infectious nodes remain in that state until they recover and gain immunity for a period that on average lasts $1/\alpha$ units of time at the end of which they rejoin the susceptible group.

To the setting above, we introduce the following isolation scheme:  If a host remains infectious for $\tau$ units of time without having recovered,  it enters a new state, $Q$ (for \textit{quarantine}), with probability $p$.  The hosts that do not enter state $Q$ at time $\tau$ remain infectious until they recover on their own. We use the terms isolation and quarantine interchangeably, and use the letter $Q$ to better distinguish it from the infectious state $I$. A host in state  $Q$  does not infect others; it
remains in this state for $\kappa$ units of time, at the end of which it is discharged and joins the recovered class. 

Figure \ref{fig:model} depicts the model states and main  parameters:
 The five states of this model are $S, E, I$, $Q$ and $R$, representing respectively  the healthy and susceptible, incubating, infectious, isolated and recovered groups within the population. 
The numbers $\sigma, \tau,\kappa>0$ and $p\in[0,1]$ are the new parameters of the model, with $\sigma$ being the incubation period, $\tau$ representing the length of time between a host's becoming infectious and  entering isolation if it is identified, and $\kappa$ the isolation time. The number  $p$ is the probability that an infectious host is identified and isolated. We refer to $p$ as the  {\it isolation probability}, and $\tau$ as the {\it identification time}. 

Table \ref{Tb1} summarizes the main parameters of the SIQ model and their meaning.

\begin{table}[htbp]
\centering
\begin{tabular}{c|l} 
\hline
parameter & meaning \\
\hline 
\hline
$\beta$ & transmission rate  \\
$\gamma$ & recovery rate \\
$\alpha$ & rate of immunity loss \\
$m$ & density of contacts \\
$p$ & probability to identify and isolate an infected individual \\ 
$\tau$ & time elapsed between infection and identification \\
$\sigma $ & incubation period \\
$\kappa$ & time spent in isolation (quarantine) \\
\hline
 \end{tabular}
  \vspace{0.2cm}
   \caption{{\small Main parameters of the SIQ model. See Table  \ref{Tb2} for specific values of parameters for given diseases.}}
   \label{Tb1}
\end{table}

{\bf Continuum limit.}  
We assume the total population is constant, and let $S(t), E(t), I(t)$, $Q(t)$ and $R(t)$ denote the fractions of susceptible, incubating, infectious, 
isolated and recovered nodes at time $t$, so that $S(t)+E(t)+I(t)+Q(t)+R(t)=1$. 

%
Assuming that links between the infectious and susceptible nodes are uncorrelated, we obtain, by moment closure, the following system of DDEs in the continuum limit as $N \rightarrow \infty$:
\begin{eqnarray}
\dot{S}(t) & = & -\beta m S(t) I(t) + \alpha R(t), \label{eq:S-dyn} \\
\dot{E}\negmedspace\left(t\right) &=& \beta m [ S(t)I(t) - S(t-\sigma) I (t - \sigma) ] \label{eq:E-dyn} \\
\dot{I}\negmedspace\left(t\right) & = & \beta mS\negmedspace\left(t - \sigma \right)I\negmedspace\left(t - \sigma \right) - \gamma I\negmedspace\left(t\right)-\beta m p e^{- \gamma \tau} S\negmedspace\left(t - \sigma - \tau \right)I\negmedspace\left(t-\sigma - \tau\right),\label{eq:I-dyn} \\
\dot{Q}(t) & = & \beta m p e^{- \gamma \tau} [S(t-\sigma - \tau )I(t-\sigma - \tau )  -S(t-\sigma - \tau-\kappa )I (t- \sigma -\tau-\kappa)].\label{eq:Q-dyn} \\
\dot{R}(t) & = & - \alpha R(t) +\gamma I(t)  + \beta m p e^{- \gamma \tau}S(t-\sigma - \tau-\kappa )I (t- \sigma -\tau-\kappa)].\label{eq:R-dyn}
\end{eqnarray}


Each term in the rate equation model \eqref{eq:S-dyn}--\eqref{eq:R-dyn} can be 
attributed to a specific transition  between the compartments, see Fig.~\ref{fig:model}. For instance, the terms 
$\beta m S(t) I(t)$ in Eqs.~\eqref{eq:S-dyn} and \eqref{eq:E-dyn} correspond to the transition from $S$ (susceptible) to $E$ 
(incubating). It is proportional to the product $S(t)I(t)$ that is, the contact probability of an
infected and a susceptible node, the transmission rate $\beta$, and the density of 
contacts $m$. 
After time $\sigma$, these incubating nodes undergo transition from $E$ (incubating) to $I$ (infectious), which is reflected in the terms $\beta m S(t-\sigma)I(t-\sigma)$.
The infectious nodes recover with the rate $\gamma$, therefore, out of the 
$\beta m S(t-\sigma-\tau) I(t-\sigma-\tau)$ nodes entering $I$ at time $t-\tau$, 
only the fraction  
$e^{-\gamma \tau}\beta m S(t-\sigma-\tau) I(t-\sigma-\tau)$ remains infectious at time 
$t$. 
These nodes are then undergo transition from $I$ (infectious) to $Q$ (isolated) with
probability $p$ corresponding to the terms $\beta m p e^{-\gamma \tau} S(t-\sigma-\tau) I(t-\sigma-\tau)$ in Eqs.~\eqref{eq:I-dyn} and \eqref{eq:Q-dyn}. The derivation of system \eqref{eq:S-dyn}--\eqref{eq:R-dyn} is analogous to that in \cite{Ruschel2018}
%
where we provided a detailed derivation for a simpler model. 
%
The results below pertain to the continuum limit system \eqref{eq:S-dyn}--\eqref{eq:R-dyn}.


\subsection*{Disease-free states and minimum response needed to squash a small outbreak}  
Without isolation and assuming everyone is susceptible, 
whether or not an outbreak will spread is determined by the disease reproductive number 
\[
r = \frac{\beta m }{\gamma}\ ,
\] 
that is, an infection spreads if and only if $r>1$ \cite{Brauer2012}.  We are interested in the case $r>1$, so that if no measures are taken the infection  will spread and reach an endemic state.  

Our first result is that when the isolation probability is $p$ and identification time is $\tau$,  the {\it effective disease reproductive number} $r_\varepsilon$ becomes
\[
r_\varepsilon =  r(1-\varepsilon) \qquad \mbox{where} \qquad 
\varepsilon:=pe^{-\gamma \tau}\ .
\]
The number $\varepsilon$ can be thought of as the {\it response capability}
of the affected community. Upon
the sudden appearance of a small group of infected individuals, the incipient epidemic
is squashed if and only if $r(1-\varepsilon)<1$, i.e., the isolation protocol as described
above reduces  the disease reproductive number by $r\varepsilon$. 
This is under the assumption that the outbreak starts near 
the disease-free equilibrium $(S,E,I,Q,R) \equiv (1,0,0,0,0)$. The case where
a  fraction of individuals is immune to begin with will be discussed later.

\noindent  We now consider separately the two components in the notion of response capability
$\varepsilon$.

{\it Minimum isolation probability.} An outbreak can be prevented only if $p> p_c$ where 
\[
p_{c}=1-\frac{1}{r}\ .
\] 
The number $p_c$ is a theoretical minimum: to have a chance to quell the outbreak, 
health authorities must have the ability to identify -- and the facilities to isolate 
-- a fraction $p_c$ of the infectious population {\it immediately} after an individual becomes infectious.

{\it Critical identification time.} Possessing the ability to isolate individuals with probability   $p > p_c$ alone  is not enough; one must be able to identify infectious hosts with sufficient speed. We prove that for each 
$p>p_c$, there is a critical identification time 
\[
\tau_{c} = \frac{1}{\gamma} \ln \frac{p}{p_c}
\]
\noindent
so that the infection dies out if $\tau<\tau_{c}$.  

Observe that in terms of what it takes to squash an outbreak, neither $\sigma$, the incubation period, nor 
$\alpha$, the rate of recovery, play a role. 
The only requirement on $\kappa$ is that it should be sufficiently 
long for the individual in isolation to recover. 
It is strictly the capability to quickly identify a large enough fraction 
of the infectious and to isolate it from the general population.

\begin{figure}
\centering
\includegraphics[width=1\linewidth]{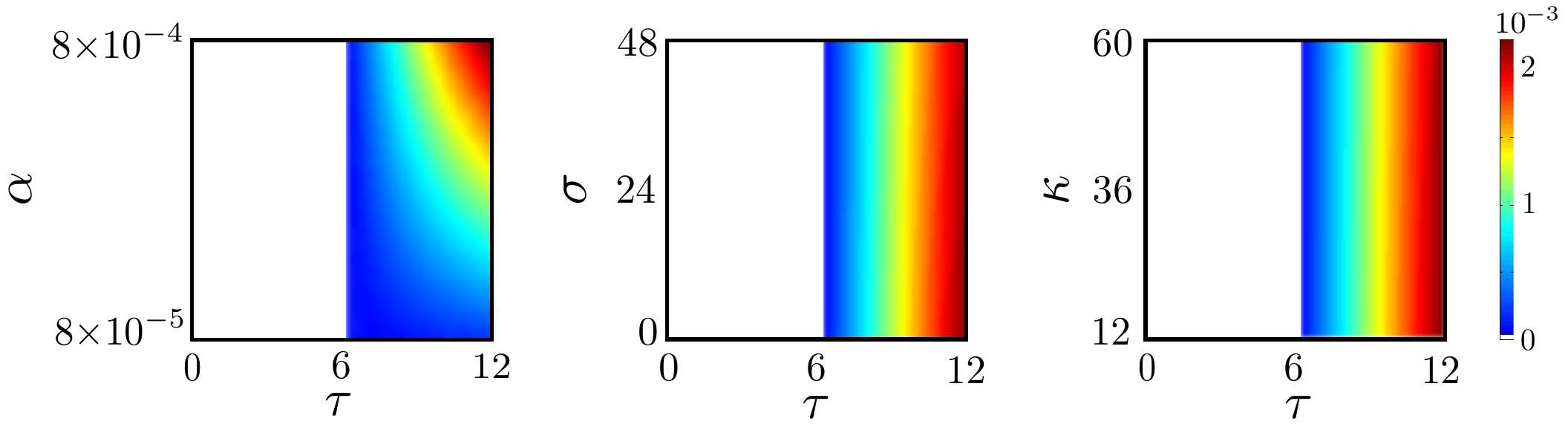}
\caption{\label{Critical_time} 
Critical identification time $\tau_c$ is independent incubation $\sigma$, isolation time $\kappa$ and  rate of immunity  loss $\alpha$.  Colors represent the asymptotic  equilibrium value of infectious individuals $I_{\rm eq}$. We consider $r = 2.5$ and $1/\gamma=12$ (as for the 2014 Ebola outbreak in Sierra Leone) with full identification probability $p=1$. Our predicted critical time is $\tau_c \approx 6.12$ days  independent of $\kappa$, $\sigma$ and $\alpha$, in agreement with simulations. For $\tau>\tau_c$, the disease-free stable state gives place to an endemic state. Left panel: $\alpha$ versus $\tau$ corresponds to the fixed values $\sigma=2.4$ days and $\kappa=60$. Center panel: $\sigma$ versus $\tau$ corresponds to the fixed values $\kappa=60$ days and $\alpha=1/1200$. Right panel: $\kappa$ versus $\tau$ corresponds to the fixed values $\sigma=2.4$ days and $\alpha=1/1200$.}
\end{figure}

Figure \ref{Critical_time} shows the results of simulations based on the  set of delay differential equations above. For illustration, we consider $r=2.5$ and $1/\gamma=12$ as in the 2014 Ebola outbreak in Sierra Leone  (see Table \ref{Tb2}). We also consider full identification probability  $p=1$. Then,  starting from a small number of infected nodes, one observes two distinct asymptotic regimes: the system tends to a disease-free equilibrium for $\tau < \tau_c \approx 6.12$ as we have shown; otherwise it tends to an endemic state. 
In the left panel,  we considered an incubation time $\sigma = 2.4$ days and an isolation time $\kappa = 60$ days, in the middle panel an isolation 
time $\kappa = 60$ days and $\alpha  = 1/1200$. Finally, in the right panel, $\alpha = 1/1200$ and $\sigma = 2.4$.  The demarkation between the two asymptotic regimes is consistent with our analytical result that critical identification time is independent of incubation time $\sigma$, isolation time $\kappa$ and the rate of immunity acquisition $\alpha$. Observe though
that for $\tau> \tau_c$ the asymptotic states do depend on these values.

\medskip 
{\it Predicting critical isolation probabilities and identification times from data}. The  critical isolation probability $p_c$ and identification time $\tau_{c} = \tau_c(p,r)$ are readily computed once the disease reproductive number $r$ is known.  Table \ref{Tb2} shows $p_c$ and $\tau_c$ for $p=0.8$, that is, $\tau_c=\tau_c(0.8,r)$, for several diseases.  
\begin{table}[htbp]
\centering
\begin{tabular}{l|ccccc} 
\hline
& $r$ & $\sigma$ & $1/\gamma$ & $p_c$ & $\tau_c$ \\
\hline
\hline
H1N1 2016 [Brazil]  &  1.7 & 4 & 7.0 & 0.41 & 4.7 \\
Ebola 2014 [Guin./Lib.]  \cite{Althaus2014} & 1.5 & 20 & 12.0 & 0.33 & 10.5\\
Ebola 2014 [Sierra Leone] \cite{Muller2015} & 2.5 & 20 & 12.0 & 0.6 & 3.5\\
Spanish Flu 1917 \cite{Muller2015}  & 2 & 4 & 7.0 &   0.5 & 3.3 \\
Influenza A  \cite{Muller2015} & 1.54 & 0.23 & 3.0 & 0.35 & 1.0 \\
Hepatitis A \cite{Peak2017} & 2.25 & 29.1 & 13.4 & 0.56 & 4.89 \\
SARS \cite{Peak2017} & 2.90 & 11.8 & 21.6 & 0.66 & 4.31 \\
Pertussis \cite{Peak2017} & 4.75 & 7.00 & 68.5 & 0.79 & 0.91 \\
Smallpox \cite{Peak2017} & 4.75 & 11.8 & 17.0 & 0.79 & 0.26 \\ 
\hline
 \end{tabular}
 \vspace{0.2cm}
   \caption{{\small Critical response capability $p_c$ and critical identification time $\tau_c$ (in days)  for various diseases with basic reproductive number $r$, and  $\sigma$ as well as $1/\gamma$ (in days). The critical $\tau_c$ is calculated  assuming that $80\%$ of infected individuals are identified and isolated. 
}}
\label{Tb2}
\end{table}

As is evident from the data shown, even with the ability to identify $80\%$ of infectious individuals, which requires considerable resources,  the critical identification time  $\tau_c$ can be as short as $3$ days for severe  outbreaks such as the Spanish Flu and the Ebola in Sierra Leone.  That is to say, to prevent such an epidemic from spiraling out of control,  infected individuals must be identified  and isolated within $3$ days of the end of their incubation period, which is typically when symptoms begin to appear and not always easy to recognize. 

While the critical isolation probability $p_c$ depends solely on the basic reproductive number $r$, the critical  delay depends on both $r$ and $\gamma$, so diseases having the same $r$ such as  Pertussis and  Smallpox can have quite distinct $\tau_c$. Recall that $r=\beta m/\gamma$. Among diseases with the same $r$, for  the same value of $p>p_c$, our formula for $\tau_c$ implies that $\tau_c(p) \propto (\beta m)^{-1}$;  that is to say, the delay one can afford is inversely proportional to the transmission rate $\beta$ of the disease for $m$ fixed, and inversely proportional to the density of links $m$ for fixed $\beta$. 

\medskip \noindent 
{\bf Immunity at the time of an outbreak.} The discussion above assumed
that at time $0$, when the infection first appears, one is near the disease-free equilibrium 
$(S,E,I,Q,R) \equiv (1,0,0,0,0)$; in particular, everyone is susceptible. We now consider 
the case where a fraction of the population has immunity at time $0$, i.e., $R(0)>0$,
possibly from a previous exposure or from vaccination. Assuming $\alpha \ll 1$, and that 
the dynamics that determine whether or not the outbreak can be contained is relatively fast, 
the situation can be approximated by one in which $R(t) \equiv R(0)$ for the time period 
under consideration. Under these conditions, our results for the case $R(0)=0$ together 
with a rescaling of the total population that participate in the dynamics yields the result 
that the outbreak is contained if and only if
$$
r(1-\varepsilon)(1-R(0)) < 1, \qquad \mbox{equivalently,} \qquad \varepsilon > 1- \frac{1}{r(1-R(0))}\ .
$$
That is, the effective disease reproductive number is scaled down by a factor of $1-R(0)$ as
expected. Having a fraction of the population with immunity allows greater leeway
in terms of minimum isolation probability and critical delay can be computed explicitly as before.


Of major concern is what happens if  the response capability is inadequate, that is, if the isolation probability falls below the required minimum, or if the required identification time is not met. Our next result addresses this scenario. 

\subsection*{Prediction of endemic states}\label{ssec:Endemic} 
\begin{figure}
\centering
\includegraphics[width=0.8\linewidth]{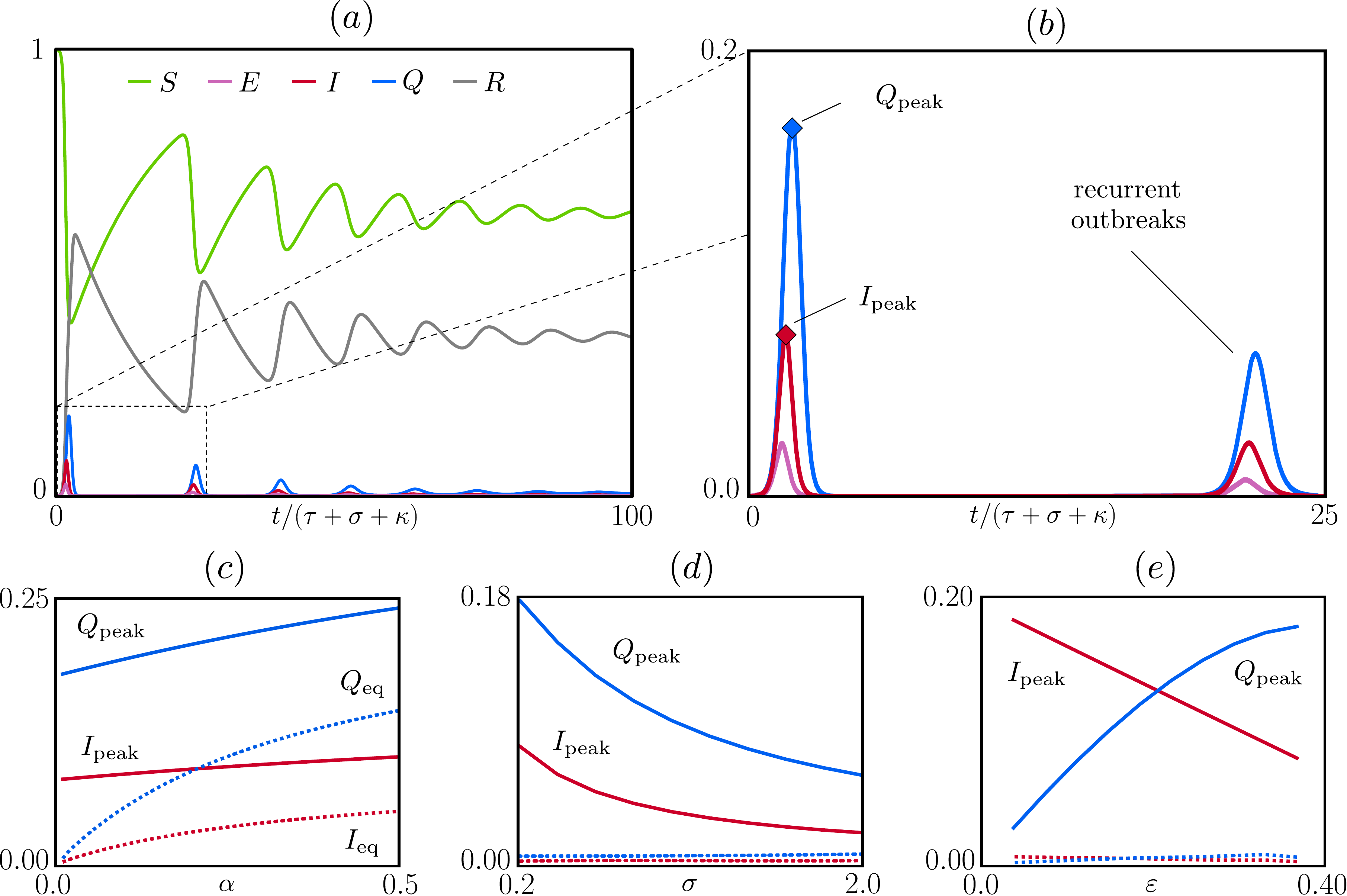}%
\caption{\label{DampedOscillators} 
Transient dynamics. (a) Time evolution of epidemic from outbreak to endemic state when
isolation protocol fails. (b) Zoom-in of (a) to show events in the first two cycles. These two panels
show the solution starting from initial condition $(S(t),E(t),I(t),Q(t),R(t))=(1,0,0,0,0)$ for $t<0$ and $(S(0),E(0),I(0),Q(0),R(0))=(0.99,0,0.01,0,0)$ corresponding to a small initial infection, with
parameters $r=2.5$, $\alpha=1/1200$, $\gamma=1/12$, $\sigma=2.4$, $\kappa=60$, 
$p=1$ and $\tau=12$. Panels (c)-(e) show the dependence of 
$I_{\rm peak}$, $Q_{\rm peak}$ (solid lines) and endemic equilibrium values $I_{\rm eq}$, 
$Q_{\rm eq}$ (dashed lines) on $\alpha$, $\sigma$, and $\varepsilon$ respectively. 
Unspecified parameters are as in (a) and (b).
}
\end{figure}

When the response capability as defined above is inadequate,  the infection will persist. That is, for parameters $p, \tau, \kappa$ and $\sigma$ such that $pe^{-\gamma \tau} > 1-\frac{1}{r}$, starting from the sudden appearance of a small group of infectious individuals, i.e., from an initial condition  near $0<I \ll 1$, 
%
$S=1-I$ and $E=Q=R=0$, 
%
the $I$-component of the solution will grow; see {\bf Methods} for more detail.  
For such initial conditions, we determine that the solution tends to an endemic equilibrium
\begin{eqnarray}
S_{\rm eq} & = & \frac{1}{r_\varepsilon} \ = \ \frac{1}{r(1-\varepsilon)}\label{eq:Seq}\\ 
I_{\rm eq} & = & \frac{ (1 -  \varepsilon)(1 - S_{\rm eq})}{ \gamma\alpha^{-1} +   \gamma\sigma  + \gamma \kappa  \varepsilon +  1 - \varepsilon} \label{eq:Ieq} \\
E_{\rm eq} & = & \frac{\gamma \sigma}{1-\varepsilon} I_{\rm eq}\label{eq:Eeq}\\
Q_{\rm eq} & = & \frac{\gamma \kappa  \varepsilon}{1-\varepsilon} I_{\rm eq} \label{eq:Qeq}\\
R_{\rm eq} & = & \frac{\gamma  \alpha^{-1}}{1-\varepsilon} I_{\rm eq}. 
\label{eq:Req}
\end{eqnarray}
A precise statement is as follows: Given $p, \tau, \kappa, \sigma$  and an initial condition as above, we prove analytically that there is a unique endemic equilibrium to which the solution can converge, and that this endemic equilibrium tends to 
%
$(S_{\rm eq}, E_{\rm eq}, I_{\rm eq}, Q_{\rm eq}, R_{\rm eq})$ 
given by the formulas above as the initial condition tends to $(1,0,0,0,0)$. 
The fact that orbits starting from near $(1,0,0,0,0)$ will approach an endemic equilibrium is difficult to prove; this part of our prediction is supported by numerical simulations. 
	Figure \ref{DampedOscillators} shows the behavior of such solutions tending to their respective endemic equilibrium, with $I >0$.
%

Observe that $\varepsilon$, the response capability, not only determines whether or not a small outbreak can be contained, it influences strongly the eventual endemic state when the outbreak happens:  $S_{\rm eq}$, the fraction of the population that remains healthy in the eventual endemic steady state, is inversely proportional to $(1-\varepsilon)$, and the fraction $I_{\rm eq}$ can be rewritten as
$$
I_{\rm eq} \ =  \ \frac{\frac1r  (r_\varepsilon - 1)}{\gamma\alpha^{-1}+\gamma\sigma+\gamma\kappa\varepsilon+1-\varepsilon} \ = \
\frac{(1-\frac{1}{r})-\varepsilon}{\gamma\alpha^{-1}+\gamma\sigma+\gamma\kappa\varepsilon+1-\varepsilon}\ .
$$
Observe in the numerators of the quantities above that $r_\varepsilon=1$, 
equivalently $1-\frac{1}{r}=\varepsilon$, is the tipping point
on one side of which the initial outbreak is contained and on the other the system tends
to an endemic
equilibrium with $I_{\rm eq}>0$. While the parameters $\sigma$, $\kappa$ and $\alpha$ 
appear in the formula for $I_{\rm eq}$, hence in $E_{\rm eq}, Q_{\rm eq}$ and $R_{\rm eq}$
as can be seen in Fig. 2, we observe that the dependence on $\sigma$ and $\kappa$ is weak
when $\alpha \ll 1$, that is, when acquired immunity lasts for a significantly longer time period 
than incubation, recovery or isolation, which is the case for most diseases.
Assuming that, we have 
\[
I_{\rm eq} = \frac{\alpha}{\gamma} \left( \left(1-\frac1r \right) - \varepsilon \right) + O(\alpha^2)\ .
\]
Here the effect of $\varepsilon$ is particularly clear: Since $1-\frac1r$ is the number
$\varepsilon$ must beat in order to contain the initial outbreak, it can be seen 
as the {\it minimal response capability} required to prevent a full-blown epidemic. 
We therefore have the interpretation that
the fraction of infectious in the endemic state is proportional to how far short the actual 
response capability $\varepsilon$ is from the minimum response needed.

\medskip \noindent
{\bf Infection cycles between outbreak and endemic equilibrium}.
Here we consider the situation where an outbreak occurs at a time no one 
in the population has immunity, such as after the disease has vanished for a couple of 
generations and suddenly returns, and the response is inadequate for curbing the initial outbreak. 
Fig.~\ref{DampedOscillators}(a),(b) depict the time evolution of the epidemic from outbreak 
to near-endemic equilibrium assuming $p, \tau$, and $\kappa$ 
remain fixed throughout. 
It is evident that the infection occurs in waves, 
or cycles, at least initially, with the size of the infection decreasing with each cycle leveling
off as it approaches endemic state. It is also quite evident that even as $I_{\rm eq}$ is small in 
the endemic state, the first wave of infection is orders of magnitude worse. 

The oscillatory behavior seen is known to occur in the SIRS model \cite{Brauer2012}
%
%
here it occurs for the same reason: as a large enough fraction of the population 
acquires immunity, and immunity is lost at a slow rate, the size of the susceptible fraction of the 
population decreases causing the rate of infection, which is proportional to $S(t) I(t)$,
 to slow down -- only to flare up again when 
the R-part of the population rejoins the S-part. What is different here is 
that we have assumed a constant response $\varepsilon$, so that for small $I(t)$,
the situation is similar to that discussed in the paragraph on $R(0)>0$ in the first part of {\bf Results}.
In particular, if $S(t)$ is such that $r(1-\varepsilon)S(t)<1$, the infection will temporarily abate,
to flare up again when $S(t)$ is large enough to bring this quantity above $1$.
Notice that the $x$-axis in Fig 3(a) is in units $t/(\tau+\sigma+\kappa)$, so that the timescales
for the cycles of infection are significantly longer than any of these quantities. Rather it is indexed
to $\alpha$ and is affected by
the fraction of individuals infected during the course of one cycle. A smaller
group is infected in the second cycle than the first because a fraction of the population is immune
at the time of the second outbreak.

We introduce the quantity $I_{\rm peak}=\max_{t>0}  I(t)$, the maximal fraction of infected individuals 
that is reached during the transient to $I_{\rm eq}$; $Q_{\rm peak}$ is defined analogously.
These quantities are of great relevance from the practical standpoint for health authorities 
involved in the planning of facilities and care for patients during isolation. To get our hands on
these quantities, we consider first the limiting case where $\alpha=\sigma=\varepsilon=0$.
In this case, our model reduces to the classical SIR model, for which we obtain
\[
I_{\rm peak}^* = 1-\frac1r - \frac{1}{r} \ln r \ ,
\]
so that $I_{\rm peak}^*$ can never be larger than $1-\frac1r$, although a direct connection between $I_{peak}^{\ast}$ and $I_{peak}$ cannot be made.
Fig 3(c),(d),(e) describe how $I_{\rm peak}$ and $Q_{\rm peak}$ vary as we permit $\alpha, \sigma$
and $\varepsilon$ to become positive. From Fig 3(c), we see that while $\alpha$ has a strong
effect on $I_{\rm eq}$ and $Q_{\rm eq}$, its effect on $I_{\rm peak}$ and $Q_{\rm peak}$ are 
relatively small, because immunity loss occurs on a much longer timescale than the time it takes
to reach $I_{\rm peak}$ or $Q_{\rm peak}$. Fig 3(d) shows that both $I_{\rm peak}$ and 
$Q_{\rm peak}$ are smaller for diseases with longer incubation periods. This presumably is 
due to the fact that a long incubation slows down the rate of infection, while recovery occurs 
at the same rate. Fig 3(e) shows that $I_{\rm peak}$ decreases as the response is stronger as
expected. The dependence of $Q_{\rm peak}$ on $\varepsilon$ is more complicated: In the figure, 
$\tau$ is fixed and $p$ is increased, i.e.,
a larger fraction of the infectious is isolated and $Q_{\rm peak}$ increases with $\varepsilon$
for this interval of $\varepsilon$, but this trend will be reversed as $\varepsilon$ continues 
to increase and $I_{\rm peak}$ continues to drop.

\section*{Discussion}

Isolation of infected hosts disconnects the infection pathway for diseases that 
are transmitted human-to-human and should, in principle, be 100\% effective, 
yet it has not always 
stopped incipient outbreaks from burgeoning into full-blown epidemics. In this paper we called
attention to the fact that a major reason may be ``too little, too late", meaning the identification
of infected hosts may not be sufficiently thorough and are too often subject to delays.
We sought a mathematical model that could offer guidance on the response needed to
contain an outbreak, as well as predict the consequences when one falls short. Our aim was to develop
a general theory that is relevant rather than to carry out detailed studies of specific diseases.
Such a model is necessarily a little idealized to be amenable to (rigorous) mathematical analysis.
With these goals in mind we came up with the system of delay differential equations (1)--(5), 
the analysis of which produced the results reported here.

We have framed response capability in terms of {\it isolation probability}, representing the fraction of the infectious population that will be identified and isolated, and  {\it identification time}, the duration an infectious individual can  transmit the disease before being isolated. An isolation strategy can prevent an outbreak only if the response capability  exceeds a minimum threshold value. This was evident in the Ebola outbreak in West Africa, where the epidemic had started to escalate  but reversed course after response capability was increased through international financial support \cite{kucharski2015}. 

We have sought to build a theory from first principles.
To stress the novelty of this class of models, which focuses on isolation including  the costs of failure in its implementation, we have taken the liberty to name the class of models considered here  {\it SIQ}, ``Q'' for ``quarantine'', to distinguish it from standard SIR type models. In our model, we do not distinguish between isolation and quarantine.  We are aware of a different use of the term ``quarantine'' in the literature to refer to the isolation of individuals who may have been infected but are not yet symptomatic \cite{Siegel2007,centers2014}. 

An interesting direction of research is to consider the isolation of individuals that have been exposed but are not yet infectious, using for example contact tracing. These strategies have the effect of increasing the isolation probability, thereby leading to possibility larger tolerable implementation delays. The influence of the underlying spatial topology of the contact network here was traded off for a clear understanding of the effects of delays in isolation.
The network heterogeneity can strongly influence the dynamics \cite{Kuperman2001} and it can either  be tackled by means of heterogeneous mean field approximation \cite{Barthelemy2005, Hayashi2004}, or by a direct study of networks with adaptive change of the structure due to network rewiring caused by the isolation protocol \cite{Shaw2008}. Such approaches should enable one to clarify the role of highly connected nodes in the isolation protocol.  

\section*{Methods}

\subsection*{Dynamical Systems Framework}
Here we present a mathematical framework for the study
of the delay system \eqref{eq:S-dyn}-\eqref{eq:R-dyn} \cite{Hale1993}. Time delay implies
that this system is infinite-dimensional, i.e. its 
state can be described by the history function, where 
the variables $S,I,E,Q,R$ are prescribed on an interval 
of the length  $\sigma+\tau+\kappa$ of maximal 
time delay. More specifically, let 
$
\mathcal C:=C\left(\left[-\sigma-\tau-\kappa,0\right],\mathbb{R}^{5}\right)
$
be the Banach space of $\mathbb R^5$-valued continuous functions with the sup-norm. Given an initial function $\phi\in \mathcal C$, the solution $x(t,\phi)\in\mathbb{R}^{5},$ $t\ge0$, to the initial value problem exists and is unique \cite{Hale1993}. 
Using the notation
\[
x_t(\phi) =x(t+\theta, \phi), \qquad \theta \in [-\sigma -\tau -\kappa, 0]\ ,
\]
it is known  that $T^t: \phi \mapsto x_t(\phi)$ defines a $C^1$ semi-flow on $\mathcal C$.

The conservation of mass property, namely 
%
$S'(t)+E'(t) + I'(t)+Q'(t)+R'(t)=0$ for all $t\geq0$, implies that if $\phi=(\phi_{S},\phi_{E},\phi_{I},\phi_{Q},\phi_{R})$ and $x(t;\phi)=(S(t),E(t), I(t),Q(t),R(t))$, then $S(t)+E(t) + I(t)+Q(t) + R(t)=\phi_{S}(0)  + \phi_{E}(0) +\phi_{I}(0)+\phi_{Q}(0) + \phi_{R}(0)$ for all $t\ge0$.  To obtain biologically relevant  solutions we further restrict to the subset on which $S(t),E(t),I(t),Q(t),R(t)\ge0$.

In {\bf Results}, we considered initial conditions $\phi$ representing the sudden appearance of a small group of infectious individuals. These initial conditions should be near 
%
$(1,0,0,0,0)$, the disease-free equilibrium with constant coordinate functions. Two examples are (i) $\phi=(\phi_{S},\phi_{E},\phi_{I},\phi_{Q},\phi_{R})$ where $\phi_S=1-c, \phi_I=c$ and  $\phi_E=\phi_Q=\phi_R=0$, all constant functions with $0<c \ll 1$, and (ii) $\phi(t)=(1,0,0,0,0)$  for $t<0$ and $\phi_S(0)=1-c, \phi_I(0)=c, \phi_E(0)=\phi_Q(0)=\phi_R(0)=0$. While such a $\phi$ is not in $\mathcal C$, $T^t(\phi) \in \mathcal C$ for $t \ge \sigma + \tau +\kappa$.
%

Numerically, DDEs are solved using the method of steps: Given an initial function $\phi$ we obtain the solution $x(t,\phi)$ to the DDE in the interval $[0,\tau+\sigma+\kappa]$ as the solution of the nonautonomous Ordinary Differential Equation using a 4th order Runge-Kutta with integration step $10^{-3}$ where the initial function is a known time-dependent input. 

\subsection*{Disease-free states} 
For fixed 
%
$\alpha, \sigma, p, \tau$ and $\kappa$, 
%
the set of initial conditions $\phi$ for which all component functions are constant and $\phi_I$ is identically equal to zero defines  the ``disease-free'' equilibria.
	These comprise the two-parameter family of constant functions $(S(t),E(t),I(t),Q(t),R(t))=(S(0),E(0),0,Q(0),0)$ for all $t$, where $S(0)+E(0)+Q(0)=1$.
 To investigate the stability of a given equilibrium solution, we linearize the system to obtain the following quasi-polynomial characteristic equation 
\[
\chi(\lambda) = \lambda^2(\lambda + \alpha) \left(\lambda + \gamma - \beta m e^{- \sigma \lambda} (1 - \varepsilon e^{-\tau \lambda})S(0)\right) =0.
\]
Two zero eigenvalues correspond to directions along the manifold of equilibria. An analysis shows that all disease-free equilibria are linearly stable if $S(0) \leq 1/r(1-\varepsilon)$, and unstable otherwise. At this bifurcation value, a third eigenvalue crosses the imaginary axis with nonzero speed and gives rise to a different class of endemic (nonzero infected) equilibria. If $S(0) \leq 1/r(1-\varepsilon)$ no further bifurcations occur, so that these disease-free equilibria are linearly stable independently of the incubation time and isolation protocol. 

The analysis above is carried out for fixed 
%
$\alpha, \sigma, p, \tau$ and $\kappa$.  We now turn the question around and ask for which values of these parameters  is $(S,E,I,Q,R) \equiv (1,0,0,0,0)$ a stable equilibrium. This is how the minimum isolation probability $p_c$ and critical identification times $\tau_c(p)$ announced in {\bf Results} were deduced.
\subsection*{Endemic states}
%
In addition to the disease-free states, system (\ref{eq:S-dyn})--(\ref{eq:R-dyn}) possesses a  two-parameter family of endemic equilibria consisting of constant functions $(S(t),E(t),I(t),Q(t),R(t)) = 
( S_{\rm eq}, E(0),  I(0), Q(0), \gamma\alpha^{-1} I(0)/(1-\varepsilon) )$ with $I(0)>0$, $S_{\rm eq}=1/r_\varepsilon$ and 
\begin{equation}
\label{eq:endemic1}
S_{\rm eq} + E(0)+ Q(0)+ \left(1+ \frac{\gamma\alpha^{-1} }{(1-\varepsilon)}\right) I(0) =1,
\end{equation}
where $I(0)$ and $Q(0)$ can be considered as the free parameters.
One sees immediately that  $S_{\rm eq}$
 is not affected by the recovery rate $\alpha$, the incubation period $\sigma$ and isolation time $\kappa$. In particular, it is constant for the whole family of endemic equilibria. The linear stability analysis of these equilibria is very challenging from an analytical point of view. They are stable for reasonable values of $\kappa$, destabilizing as $\kappa$ increases. A detailed analysis will be published elsewhere. The formal limiting case $\alpha\to\infty$ (immediate waning of immunity) has been studied in detail \cite{Ruschel2018}.

The main concern of this paper is the effect of isolation on disease outcome starting from a small outbreak.  To this end, the following conserved quantities will be of major help:
\begin{eqnarray}
H_E(\phi) &=& \phi_E(0) - \beta m \int_{-\sigma}^0 \phi_S(\theta)\phi_I(\theta)\mbox{d}\theta \nonumber\\
H_Q(\phi) &=& \phi_Q(0) - \beta m \varepsilon \int_{-\sigma-\tau-\kappa}^{-\sigma-\tau} \phi_S(\theta)\phi_I(\theta)\mbox{d}\theta \nonumber
\end{eqnarray}
%
The conservation of these quantities along solutions
$\frac{d}{dt}H_{E,Q}(x_t(\phi))=0$
can be proved by direct computation. 
%
Restricting our attention to initial functions of the type
$(1,0,0,0,0)$ for
$t\in[-\tau-\sigma-\kappa,0)$ and $(1-I_{0}(0),0,I_{0}(0),0,0)$
at $t=0$ for some $I_{0}(0)$, we have 
$H_{E,Q}(x_t(\phi))=H_{E,Q}(\phi)=0$. This implies
\begin{eqnarray}
E(t) & = & \beta m\int_{t-\sigma}^{t}S(\theta)I(\theta)d\theta,\label{eq:2-1-1}\\
Q(t) & = & \beta m\varepsilon\int_{t-\tau-\sigma-\kappa}^{t-\sigma-\tau}S(\theta)I(\theta)d\theta.\label{eq:4-1-1}
\end{eqnarray}
Equations \eqref{eq:2-1-1} and \eqref{eq:4-1-1} can be used to find the corresponding endemic equilibrium $(S_{\rm eq},E_{\rm eq},I_{\rm eq},Q_{\rm eq},R_{\rm eq})$ reached from the 
considered initial function. More specifically, we 
determine $E_{\rm eq}=\beta m \sigma S_{\rm eq} I_{\rm eq}$ and $Q_{\rm eq}=\beta m \varepsilon \kappa S_{\rm eq} I_{\rm eq}$. 
Inserting these values in Eq.~\eqref{eq:endemic1}, we obtain 
$I_{\rm eq}$, and, hence, the unique endemic equilibrium \eqref{eq:Seq}--\eqref{eq:Req}.

{\bf Acknowledgments}: This paper was developed within the scope of the IRTG 1740/ TRP 2015/50122-0, funded by the DFG/FAPESP. L-S.Y. was partially supported by NSF Grant DMS-1363161. T.P. was partially supported by  FAPESP grant 2013/07375-0 and Serrapilheira grant G-1708-16124. T.P. and S.R. thank the Courant Institute for the hospitality. 

\noindent
{\bf Author Contributions}: All authors carried out the research. L-S.Y. and T.P. wrote the main body of the manuscript. All authors wrote and edited the manuscript. S.R. and T.P. conducted numerical experiments. L-S.Y. designed Figure 1, T.P. prepared Figures 1,2 and S.R. and S.Y. prepared Figure 3.

\noindent
{\bf Additional Information} \\
\noindent
{\bf Competing interests}: The authors declare no competing interests.

\end{document}